# Robust Physical Encryption and Unclonable Object Identification in Classical Optical Networks using Standard Integrated Photonic Components

Jack A. Smith and Michael J. Strain

Institute of Photonics, Dept. of Physics, 99 George St., Technology and Innovation Centre, University of Strathclyde, Glasgow G1 1RD, UK

Corresponding author(s): jack.smith@strath.ac.uk and michael.strain@strath.ac.uk

**Abstract**

Spectral complexity – a useful resource in physical device identification, disorder-enhanced spectroscopy, and machine learning – is often achieved in chip-scale devices at the expense of propagation loss, scalability, or reconfigurability. In this work, we demonstrate that device specific spectral complexity can be achieved using completely standardized photonic building blocks. Using a waveguide Mach-Zehnder interferometer internally loaded with two sets of non-concentric dual ring resonators, we demonstrate the generation of unclonable keys for one-time pad encryption which can be reconfigured on the fly by applying small voltages to on-chip thermo-optic elements. With this method, we access a keyspace larger than 12 Tb for a single device with simple, single-mode waveguide input and output coupling. Using two devices at either end of a communication channel, we show that an eavesdropper tapping the channel fibre link would be unable to recover the same spectrum measured at either end of the link, providing physical encryption for key distribution. Furthermore, being purely classical, this form of secure communications does not require quantum photonic sources or detectors, and can therefore be easily integrated into pre-existing telecommunication architectures.

## 1. Introduction

### 1.1 One-Way Functions and One-Time Pads

Today's highly interconnected digital society owes its security to mathematical one-way functions. For example, it is computationally simple to multiply two prime numbers together, but prohibitively time consuming to factorize into two unknown primes a given large integer. Problems such as these form the basis of modern cryptography, where secrecy can be assumed only so long as there exists no algorithm capable of solving these problems on a classical computer in a reasonable time.

However, a sufficiently powerful quantum computer running Shor's algorithm could find these primes in polynomial time, rendering today's asymmetric public key encryption insecure[1]. With this in mind, it is conceivable that sensitive ciphertext is being stored now to be decrypted in the future by a hypothesized cryptographically-relevant quantum computer. The threat of such "harvest now, decrypt later" attacks has motivated the development of post-quantum encryption algorithms, which are more resistant to decryption by quantum computers[2], [3], [4], [5]. However, such post-quantum encryption algorithms are still based on mathematical problems whose solutions at first seem intractable, but may in fact exist and even be tractable for a single classical computer[6].

Besides mathematical one-way functions, there has been great interest in *physical* one-way functions, which are also known as physically unclonable functions (PUF)[7], [8]. Here, security is a result of the entropy intrinsic to the physical object or token itself. The PUF can take many forms, but its key characteristics are that for a given *challenge* it will repeatably offer a well-defined *response*. These responses cannot be predicted from known challenges, and they cannot be spoofed even with complete knowledge of the physical system inside the black box. For example, a speckle pattern (the response) can be generated by shining a laser of a certain wavelength (the challenge) at a random scattering token (the PUF)[9]. The stochastic nature of the distribution of scatterers in the token is not physically reproducible, being unique to that PUF only and thus guaranteeing the authenticity of the speckle pattern. For cryptographical applications, random keys can be generated from these PUF responses and used for one-time pad encryption (OTP), whereby ciphertext is usually generated by taking the bitwise XOR operation between the pre-shared PUF key and the plaintext message. Shannon showed that if an OTP key is never reused, is random, and (of-course) is kept secret, then it is completely secure[10]. Classical OTP methods offer an alternative to quantum key distribution (QKD) techniques, which in ideal circumstances offer physical security underpinned by the no-cloning theorem. Classical methods have an advantage over quantum encryption schemes in the ability to employ traditional emitters and detectors operating at reasonable transmission power levels, mitigating the effects of loss in communication channels that are a substantial challenge in quantum schemes[11]. The major challenge in classical systems is the distribution of OTP keys on public channels that are robust to eavesdropping attacks[12].

**1.2 Integrated Optical PUFs**

Macro/mesoscale optical PUFs like the scattering token example above can be difficult to integrate with fibre transmission channels, since the resulting disorder is often imprinted in the spatiotemporal profile of speckle patterns or complex optical modes, rather than being encoded in the spectral outputs of conventional single-mode waveguides. Conversely, photonic integrated circuits (PICs) can be more easily integrated with fibre, but are not readily thought of as having spectrally complex outputs and disordered structures. Indeed, the vast majority of the technological applications of PICs are underpinned by the well-behaved and well-understood transmission functions of standard building blocks, such as directional couplers, ring resonators (RRs), and Mach-Zehnder interferometers (MZIs). Thus, when targeting an application such as OTP, one may be inclined to entirely forgo this standard toolkit of photonic integrated devices, and instead opt for a structure with minimal symmetry[13], [14] or high numbers of embedded scatterers[12].

To this end, integrated devices employing Aubry-André analyticity breaking [13], [14], [15] or fully chaotic wave trajectories [12], [16] have so far been explored, all utilising SOI waveguides. In order to resolve the sub-wavelength features that are key to the functioning of these devices, high-resolution electron beam lithography is required. In single device acquisitions, key lengths of $10^3$ [12] to $10^4$ [15] have been demonstrated. Full keyspaces, accessed through mechanical reconfiguration of the input fibre position in the former case and Joule heating of devices in the latter, are of the order of 0.1 Tb and 20 Mb respectively. Since OTP encryption requires keys at least as long as the plaintext message, long keys are highly desirable. Furthermore, since keys may not be reused, devices and their spectral outputs must also be in some way reconfigurable. The main challenge that this work addresses is how to effectively synthesize these two requirements in a scalable fashion; namely, how to translate a very large space of available keys, reconfigurable on-the-fly, into the outputs of standard optical waveguides with low loss.

For the separate application of device authentication, it is necessary to be able to recover the same key twice under separate interrogations of one physical device. Given the exponential sensitivity to initial conditions inherent in chaotic wave dynamics, this may lead to difficulties in using the aforementioned fully chaotic PUFs for this application. Thus, devices whose input and output

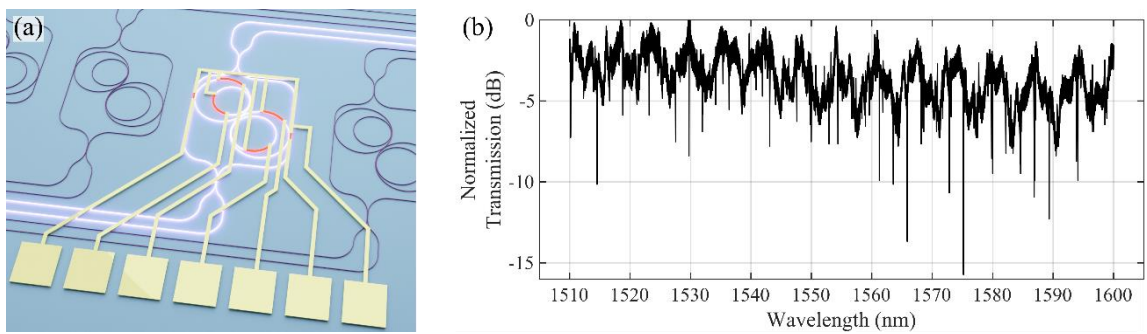

**Fig.1 (a)** Schematic representation of the device, with ring heaters labelled in red. **(b)** Example device transmission spectrum.

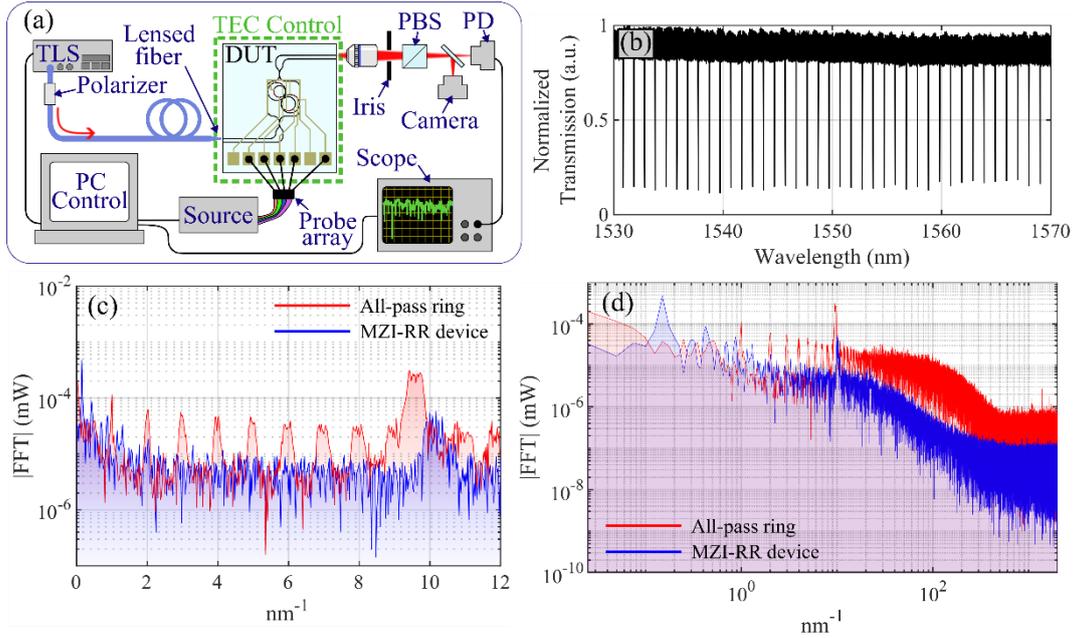

**Fig.2 (a)** Optical setup used to probe device. **(b)** Transmission spectrum of a comparison all-pass ring fabricated on the same chip. **(c)** FFT of the all-pass and MZI-RR spectra, comparing low frequency components. **(d)** Full FFT spectra of the two devices.

conditions can be more deterministically set with standard technologies have the dual use of OTP key generation and physical device authentication.

In this work, we present a silicon nitride waveguide device composed of standard low-loss photonic building blocks, but whose optical transmission spectra offer rich spectral complexity, sensitive both to fabrication variations and changes to ambient conditions during optical interrogation. When used to generate keystreams from transmission spectra for OTP encryption, the aforementioned sensitivities and complexities make it difficult to replicate or spoof the keys generated by the device, especially given OTP keys are used only once, and integrated microheaters allow rapid reconfiguration of the transmission characteristics of the device between key generation times. However, repeated interrogations of the same device under the same driving conditions produce the same keys. Furthermore, since the inputs and outputs of the device are edge-coupled SiN waveguides, keys can be extracted quickly and easily into standard fibre-optic components, greatly enhancing the compatibility of our device with pre-existing telecommunication architectures.

## 2. Designing for Irreproducibility

The device, schematically shown in Fig. 1a and hereafter referred to as *MZI-RR*, is composed of an imbalanced MZI whose arms are internally connected by two sets of two non-concentric internally coupled ring resonators. Further details can be found in supplementary note 1. All four rings have slightly different radii. Fabrication was carried out by LIGENTEC (Switzerland) as part of a multi-partner wafer run in their AN800 platform, which features an 800 nm SiN waveguide core, leading to a tightly confined mode at c-band wavelengths. The waveguide width was 1 μm and coupling gaps were 350 nm. Four integrated microheaters could be used to individually tune the rings. Heating sections are labelled in red in Fig. 1a, and an optical micrograph of the device can be seen in Fig. 6a.

The entire geometry was designed so as to promote multiple optical trajectories and resonance-assisted pathways of differing path lengths, phase accumulations, and coupling strengths. These pathways are collected in a feed-forward fashion and interfered at the MZI output directional coupler. In all, there are 7 evanescent waveguide-waveguide coupling sections in the device, offering multiple opportunities for even small device-to-device fabrication variations to alter transmission functions with respect to their nominally identical counterparts across chip copies. Furthermore, small deviations from nominal waveguide widths and heights lead to variations in the effective index of guided modes [17]; and finite difference eigenmode simulations indicate that a 1 nm reduction in waveguide width reduces the effective index of the SiN mode by 0.0002. Given the extremely low propagation losses of the SiN platform and the sharpness of resulting resonances, this effective index difference is enough to shift a resonance by several linewidths, and in a nested configuration such as ours, this will lead to significant variations in the output spectra. These two mechanisms, which both ultimately relate to the exponential function in the decay of the optical field outside of the waveguide core, are assumed to be the dominant causes of stochastic device-to-device transmission variance, and this is the basis of their being used as physically unclonable functions for OTP encryption.

### 3. Analysis of Transmission Characteristics

The setup shown in Fig. 2a, featuring a narrow-linewidth tuneable laser and a high-resolution oscilloscope, was used to test the device's optical transmission characteristics. More experimental details can be found in supplementary note 2. The following analyses are for TE polarization and the same combination of MZI in/out ports (in port 1, out port 3, see supplementary note 1). It should be noted however that the device has similar performance for TM polarization, and measurements from different port combinations, as shown in supplementary note 3.

Fig. 1b shows the transmission spectrum of a typical device, which was taken by collecting the output from port 3 while injecting the tuneable laser source into port 1. The response is characterized by a low frequency modulation whose free-spectral range (FSR) of ~6.3 nm matches the theoretically predicted value for an MZI with this geometry. We therefore attribute this modulation envelope to the MZI. There is also a higher frequency modulation which is due to the waveguide facet-to-facet reflectivity setting up a reasonably long Fabry-Perót cavity. This is common to all devices and visible in the detail of Fig. 6b. There are multiple distinct resonances of differing linewidths and extinctions, detailed in Fig. 3. We emphasize that so many lineshapes are produced by the device because the spectrum measured at the output port contains information on the multiple resonance-assisted pathways the light may have taken to get there, including contra-directional coupling, single or multiple notch passes, and single or multiple drop-port peaks coupled by the rings across the two MZI arms before being interfered at the output directional coupler of the MZI. In all, this leads to both sharp, low-loss resonances, and more broadband peak and dip modulations in the output. Details of the resonance fitting can be found in supplementary note 4.

We consider now in more detail the FFT frequency spectrum of the transmission function, and it is informative at this point to introduce a comparison with an all-pass ring resonator, also fabricated on this chip. Its transmission spectrum is shown in Fig. 2b, and the FFT of both are shown in Fig. 2c-d. At the low frequency end, it is clear to see that obvious ring structure such as a fundamental frequency

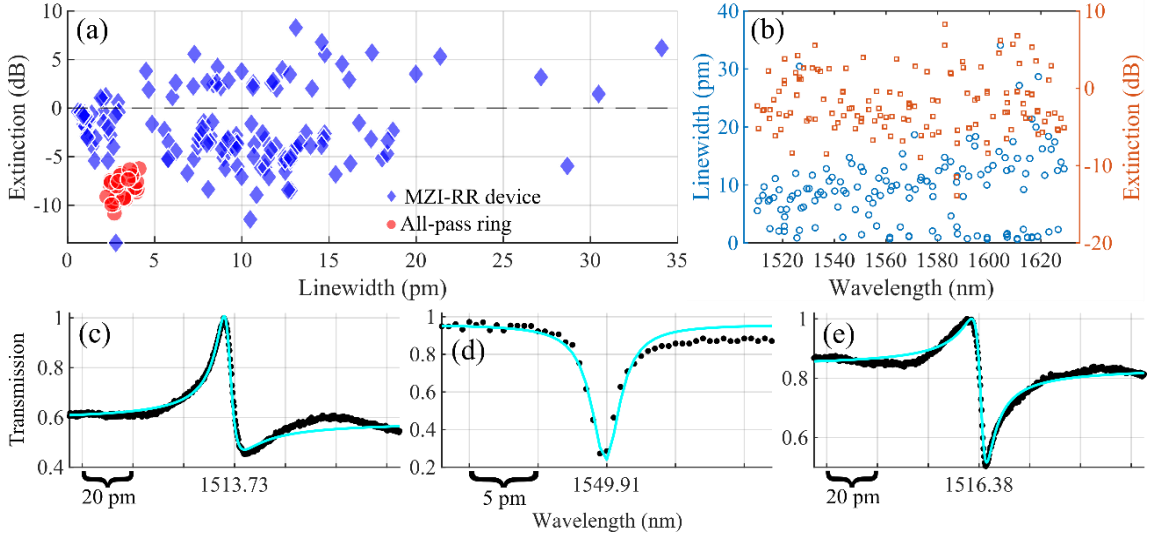

**Fig.3 (a)** Comparison of the extinctions and linewidths of the resonances produced by the MZI-RR device and comparison all-pass ring. **(b)** Resonant linewidths and extinctions of the MZI-RR device as a function of wavelength. **(c)-(e)** Resonance fits of some example lineshapes, with Q factors of **(c)** $2.30\times10^5$, **(d)** $1.25\times10^6$, **(e)** $2.70\times10^5$

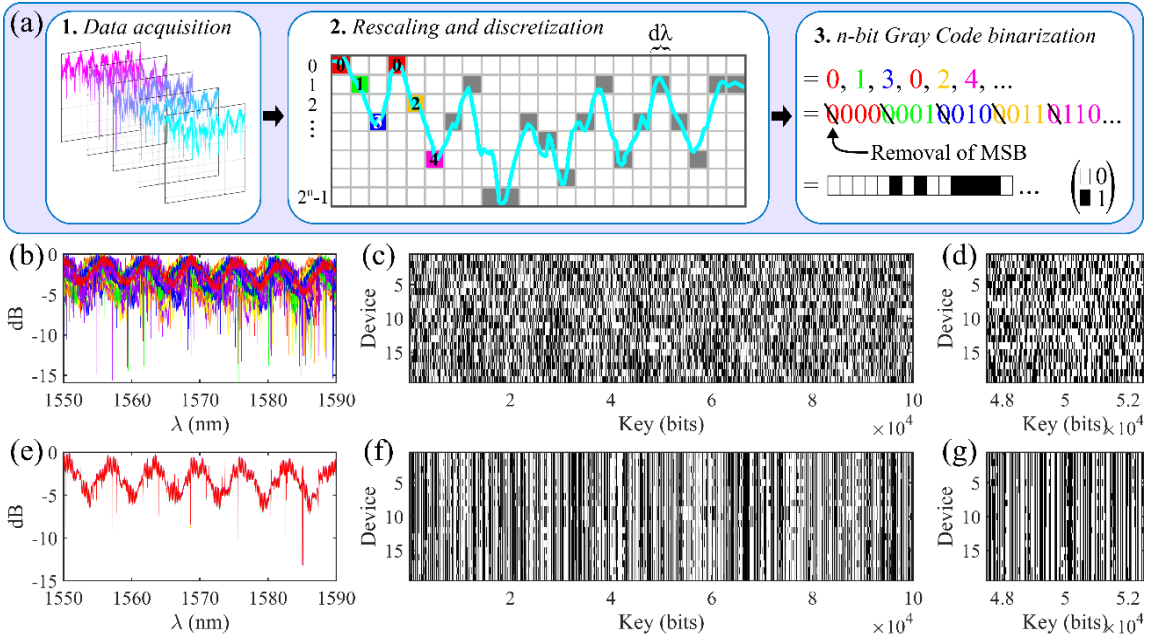

**Fig.4 (a)** Method for extracting keys from device spectra, following a method outlined in [13], and explained in detail in the methods section. **(b)** Raw normalized transmission spectra for 20 nominally identical devices. **(c)** Keys extracted from these spectra, for *n = 6 bits*, *dλ = 0.002 nm*, and most-significant bit removal. **(d)** Detail of the keys around a central 5000 bit (2 nm) window. **(e)** Raw transmission spectra for 20 repeated measurements of one device. **(f)** Keys for these spectra, with the same settings as previous. **(g)** Detail of those keys, with same key extractin settiongs as (c-d).

component and dispersive harmonics is largely absent from the MZI-RR device, which is instead dominated by a main peak corresponding to the aforementioned FSR of the MZI envelope function. Both MZI-RR and all-pass feature broad peaks at around 10 nm$^{-1}$, corresponding to the Fabry-Perót FSR. Besides these components, there is little information in the frequency analysis of the MZI-RR that could be used to construct a facsimile of its wavelength response, as could be done for the all-pass ring.

Access to the all-pass ring also allows for nonlinear least squares fits of its lineshapes, revealing distributed ring losses of just $0.19 \pm 0.03$ dB/cm. The loss for the MZI device can be inferred to be similarly low since the waveguide geometries are the same and bend radii are both above cutoff. Fig. 3 details a more in-depth analysis of the resonant lineshapes available from the MZI-RR and the comparison ring. Naturally, a ring's resonances do not vary significantly over reasonably small bandwidths, and so their extinctions and linewidths cluster around values determined by the round-trip ring loss and bus-to-ring coupling coefficient. In the MZI-RR however, there are multiple round-trips, multiple losses, and multiple effective coupling coefficients, and so the lineshapes are far more varied. Given we are also effectively collecting the drop-port of the entire system in feed-forward fashion, the resonances can also be peaks instead of dips. This information is collated in Fig. 3a-b, where Fig. 3b also shows no obvious trend in resonance characteristics with respect to wavelength. Some example spectral feature fitting using a non-linear regression method, of both Lorentzian and pseudo-Fano type features, are also shown in Fig. 3c-e.

## 4. Key Extraction Procedure and Validation of Physically Unclonable Functions

Having demonstrated the transmission characteristics of one MZI-RR device, we now detail how transmission spectra taken from multiple devices (or a single device under varying microheater voltages) can be digitized and converted into binary keys. The process is graphically shown in Fig. 4a, and follows a method laid out in Tarik, et al[13]. The full process is detailed in the Methods section. With this method it was possible to convert spectra into 1-dimensional keystreams and compare statistics including inter-key correlation and the Hamming fraction, which is simply the fraction of positions in two keys which do not share a common bit, meaning a Hamming fraction of 0 indicates identical keys and 1 corresponds to completely non-identical keys. Since in the latter case a simple bit-

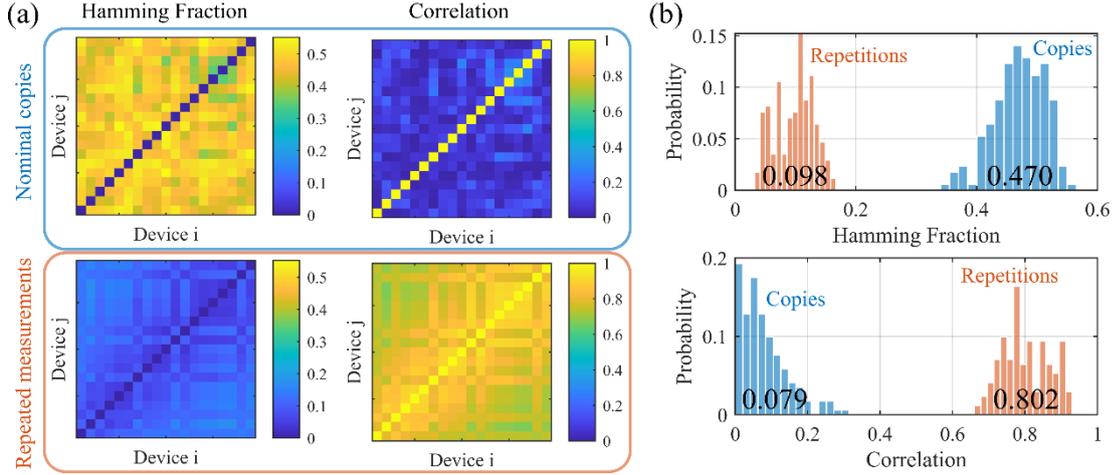

**Fig.5 (a)** Hamming fractions and correlations between all pairs of keys shown in Fig. 4, both the nominally identical copies, and repeated measurements of one copy. **(b)** Histogram data for the off-diagonal elements in (a), with mean values labelled.

wise NOT operation would recover identical keys, the ideal Hamming fraction for minimally similar keys is 0.5.

Fig. 5 compares these statistics for the two datasets shown in Fig. 4, namely measurements comparing nominally identical device keys, and repeated measurements on a single device. The auto-correlation and auto-Hamming fraction are always 1 and 0 respectively. The off-diagonal elements are of greater interest, and histograms of their measured values are displayed in Fig. 5b. We find that nominally identical devices produce measurably different keys, with a mean Hamming fraction of 0.470±0.042 and correlation of 0.079±0.064. The simple Euclidean distances between measured spectra is also detailed later in section 6. In contrast, repeated measurements of a single device present a much lower Hamming fraction of 0.098±0.032 and higher correlation of 0.802±0.065. This indicates that device-to-device variations are sufficient security against the spoofing of keys, even if an adversary has access to the device design and exact fabrication technology. The ability to reliably recover statistically similar results from one device, but dissimilar results from nominal copies of that device, is also important for verifying device authenticity. It is also worth noting the spectral resolution of the characterisation at 0.25pm, and non-repeating spectral features in the few pm range, make replication of these measured spectra by other means prohibitive with current and reasonable foreseeable future technologies due to the asymmetric difficulty of measuring and synthesising optical spectra.

## 5. Device Reconfiguration Through Electronic and Non-Linear Optical Means

We now show how the keys generated by the MZI-RR device can be reconfigured, greatly expanding the number of keys available from a single physical device. This can be achieved by applying small voltages to the micro-heaters individually tuning each ring resonator. If each combined voltage setting is treated as producing a new virtual device, the keys generated from their spectra can be compared with each other in exactly the same way as presented above. Initially, however, it is highly informative to simply apply a 5 V bias to only one ring (labelled in Fig. 6a). It is clear from Fig. 6b that tuning one ring does not result in a basic red-shifting of one resonance family only. Instead, a non-trivial effect on the output spectrum is observed. Some peaks are completely extinguished, some are shifted and their lineshapes altered, and the MZI envelope function features localized modifications to its extinction, indicating narrowband phase variations between the two MZI arms.

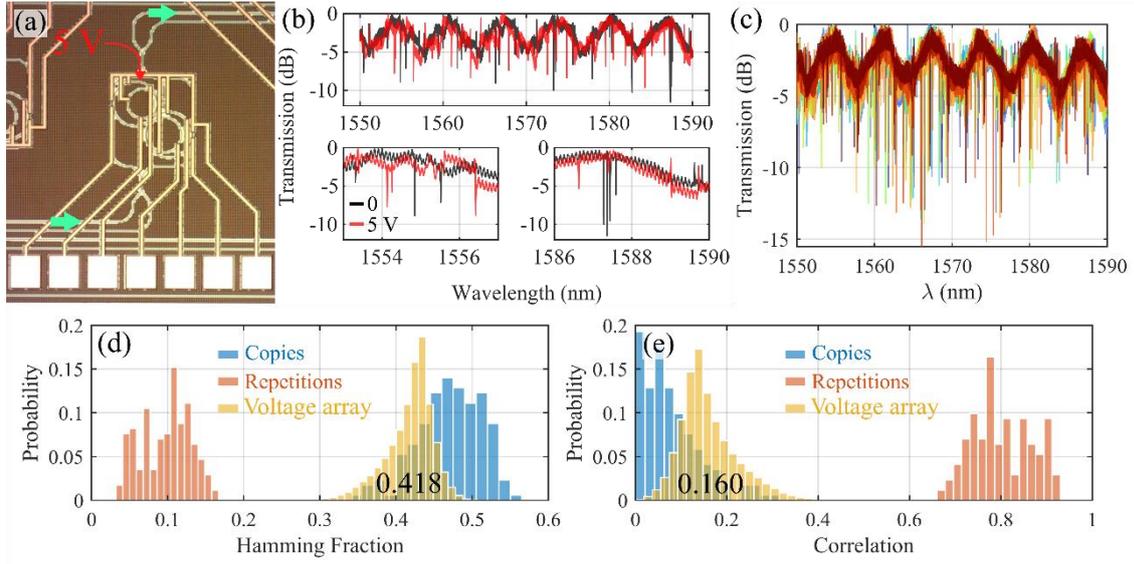

**Fig.6 (a)** Optical micrograph of the device. **(b)** Transmission spectra before and after applying 5 volts to the labelled heater. **(c)** A random selection of 14 transmission spectra from 216 repeated measurements where randomized voltages were applied to the ring heaters. **(d-e)** Off-diagonal Hamming and correlation histograms of the 216 random voltage keys, with mean values labelled, overlaid with those of Fig. 5 for comparison.

Next, a series of voltage configurations were applied to the microheaters, and keys generated for each. In all, 216 different spectra were taken in an automated fashion. First, randomly sampled voltages taken from the range 0 to 5 V were applied to the heaters, and the spectrum was measured as before. A sample of 14 normalized spectra from the 216 total is shown in Fig. 6c. The Hamming fraction and correlation statistics for all the generated keys are shown in Fig. 6d-e, which for comparison, also includes the results for the nominal copies and repetitions detailed previously. Strong overlap with the previous case of measurably dissimilar keys is found, although there is a slight reduction in performance in terms of a decreased mean Hamming fraction and increased correlation. Nonetheless, this shows that a single physical device can be repurposed to produce multiple keys post-fabrication, by utilizing the non-trivial response of the MZI-RR configuration to small voltages applied to the microheaters.

In addition to voltages applied to thermo-optic resonance shifters on-chip, the intensity-dependent optical nonlinearities of the SiN waveguides themselves may be leveraged to further increase the keyspace of a single device. Using an EDFA before the injection fibre to increase the on-chip power, the transmission spectra shown in Fig. 7a were taken. The differences between the spectra are not as rapidly varying as a function of injection power, as was the case for thermal actuators, but there are still clear and non-uniform variations in resonance position, lineshape, and extinction. Such variations in the spectra tend to be highly localized, and most probably correspond to only the most long-lived resonant pathways through the device and the thermal non-linear effect common in high quality factor resonators. Thus, although the actual difference between each spectrum may be small, the resulting lineshapes in those regions that do change remain difficult to predict from first principles, and can be used as another method to add variability to the available keys, or to authenticate the response of an individual device.

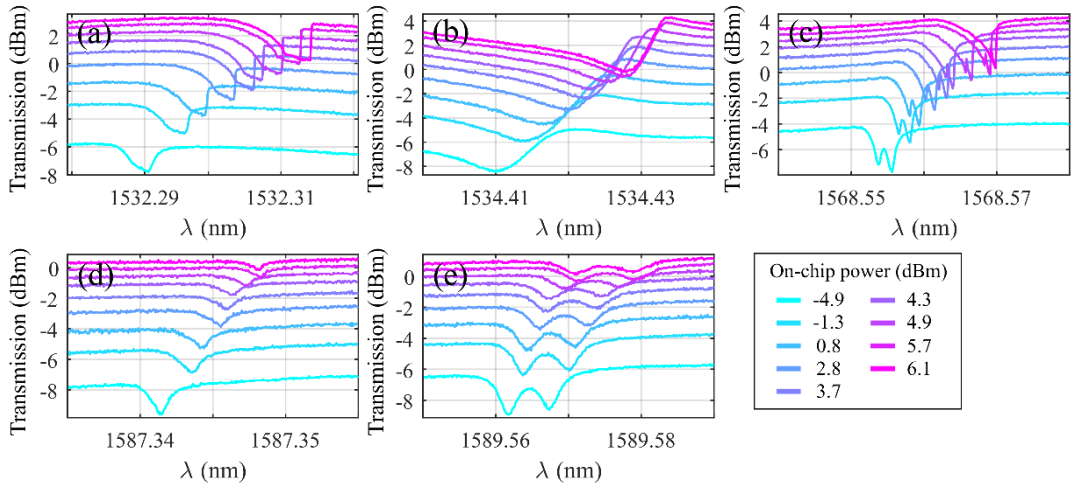

**Fig.7** Detail of resonances under increasing on-chip optical powers, featuring **(a)** apparent thermal bistabilities, **(b)** red-shifting of broader spectral features, **(c)** differential shifting of nearby or split-resonances, **(d)-(e)** red-shifting and extinction reduction.

## 6. Demonstration of a Key-Distribution Channel Robust to Eavesdropping

Finally, we investigate the feasibility of using pairs of chips to share a key securely between two parties. Two chips labelled A and B were selected randomly from the fabricated set and arranged as in Fig. 8a. They were measured in two directions: forward (AB) and backwards (BA). In either case, the fibre channels between the laser and device or photodetector and device are assumed to be secure, since these would be located at the ends of the communication link. The weak point is the fibre connecting the two devices, which may freely be tapped by an adversary intending to secretly measure the optical signals

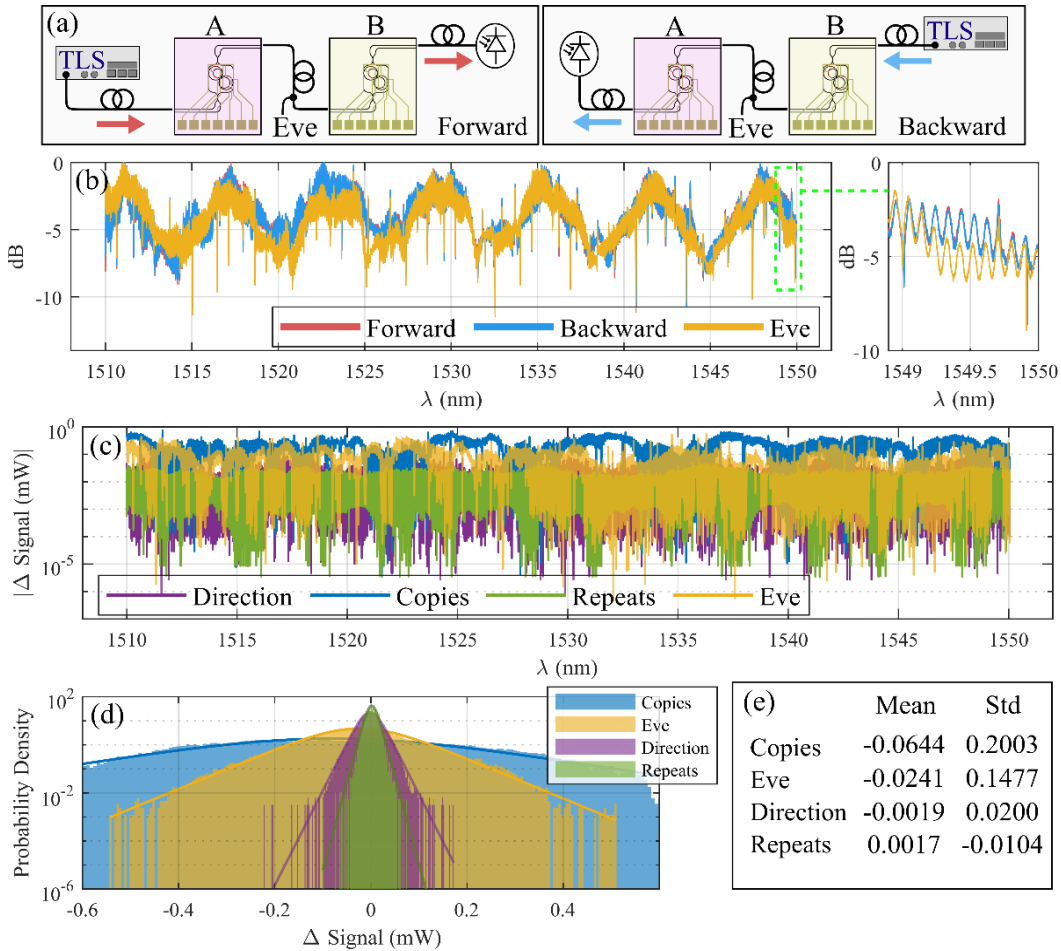

**Fig.8 (a)** Forward and backward directions in double PUF key sharing experiment. **(b)** Transmission spectra through the two chips, comparing forward (AB), backward (BA), and an eavesdropper who constructs their signal from (A×B), with inset (right). **(c)** Euclidean distances between the spectra, with the difference between two randomly selected chip copies and two repeated measurements of one chip from Fig. 5 also included. **(d)** Probability densities, with fitted probability density functions, of the differences detailed in **(c)**. **(e)** Table of statistics for each of the differences.

shared between A and B. The adversary also has complete knowledge of the measurement procedure and key extraction protocol.

First, note that as expected the difference between the forward and backward directions is comparable to the case of repeated measurements of a single device. This is apparent both in direct comparison of their respective spectra (Fig. 8b) and in the statistics of the Euclidean distances between each dataset, which are similar to those of the previously discussed repeated measurements of a single device (c.f. Fig. 5). This single-device repeated measurement can be used as the baseline for measurement similarity. If an eavesdropper were to tap the fibre link between the two devices, they would have full access to the spectra of the individual devices A and B. However, by linking two PUFs over fibre, we have effectively constructed an entirely new, distributed PUF. The phase dependent spectral response of this combined device is only accessible at the secure ends, but not from a fibre tap in the channel, which instead is only capable of measuring the signal amplitude through chips A and B individually. A reconstruction of the expected full spectrum can then be attempted by multiplying these individual spectra. The eavesdropper's reconstructed spectrum (labelled "Eve" in Fig. 8b) is able to approximate some of the sharper resonant features, though with some notable differences in extinction. More obvious is the significant difference in the slowly varying components of the spectrum, which indicates the eavesdropper's failure to capture vital phase information shared across both chips. This can be seen in the statistics of the difference between Eve's reconstructed spectrum and the legitimate combined spectrum. These differences are much more akin to the previous case of nominally identical devices which produce vastly different spectra, which have also been included in Fig. 8c,d.

Using the same key extraction procedure and settings as previously discussed, the forward (AB), backward (BA), and eavesdropped (A×B) keys can be compared. These results are shown in Fig. 9, where there is a clear differential between the legitimately shared keys and the eavesdropper's copy. If the labelled off-diagonal values are compared to those plotted in Fig. 6d,e, we find the comparison between forward and backward directions falls within the grouping of repeated measurements which sets the baseline for similarity, while the eavesdropper's copy falls within those measurements of dissimilar keys.

## 7. Discussion

The photonic integration of physically unclonable functions composed of standard components compatible with the design rules of commercial foundries represents an important advancement in classical cryptography, and in particular, offers a more scalable alternative to other methods. We have shown that the natural fabrication variations between nominally identical devices are enough to give them radically different transmission characteristics. Fabrication non-uniformity in photonic chips is a well-documented challenge in systems-level design and manufacturing [18], [19], [20], [21], [22]. For integrated photonic unclonable functions however, these stochastic fabrication variations are key to ensuring that no two devices are exactly the same, and subsequently, that their respective optical responses are unique. The complexities in exactly how these fabrication differences manifest changes in the optical outputs are what provides security against synthesizing their responses, even with full knowledge of the chip design and fabrication processes.

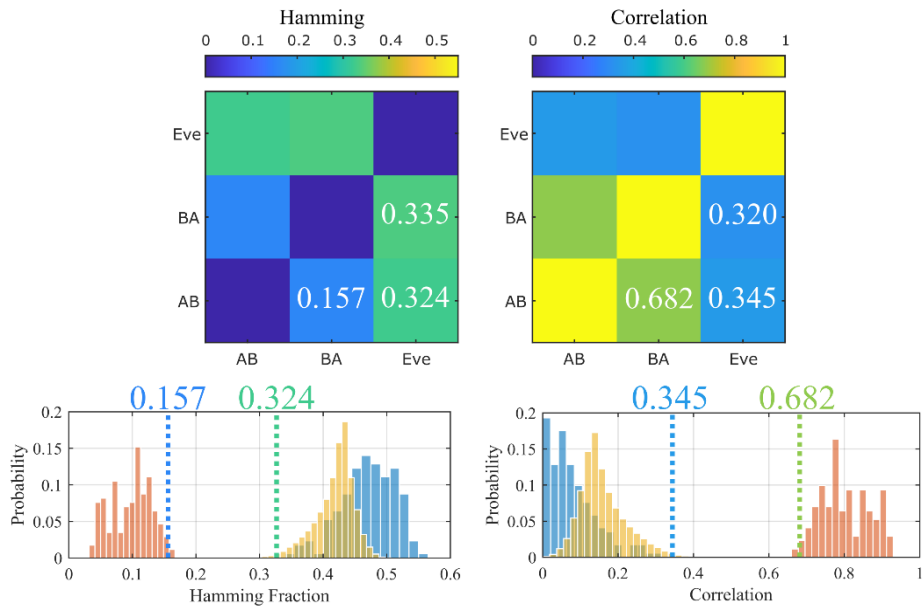

**Fig.9** Hamming fraction and correlation statistics for the keys extracted from the forward, backward, and eavesdropper reconstructed measurements.

Prior works have sought to enhance the entropic response of PUF devices by targeting geometries that specifically lead to chaotic ray trajectories[12], [16] or Aubry-André analyticity breaking[13], [14]. The work presented here is an alternative design methodolgy for integrated PUFs employing foundry-compatible, standard, low-loss, single mode SiN waveguides, as opposed to complex SOI structures. By using single mode waveguides we ensure compatibility with fibre-optic input/outputs, while also greatly limiting the possible effects of fibre-chip alignment on the ensuing spectra. Furthermore, by operating in a low-loss regime with propagation losses less than 0.19 dB/cm, we ensure that our device contains both long-lived resonances as well as short-lived ones. In a Mach-Zehnder device such as ours, which is sensitive to pathway phase differences, fabrication variations that alter these photon lifetimes will lead to a characteristic intensity response at the output. Thus our device is sensitive to fabrication variations that alter both phase and amplitude. The narrow phase-sensitive spectral features, which we have also shown to be intensity dependent (Fig .7), serve two purposes. First, they ensure localized complexity on small wavelength scales, allowing for a high sampling rate in wavelength and in turn for very long keys extracted from one transmission sweep. Second, if they have been pre-characterized, their individual responses and lineshapes can be used to validate the authenticity of that PUF token.

For OTP encryption, with a subsampled wavelength discretization of 2 pm as used above, 6 bits with most significant bit removal, and the full sweep span of the laser (90 nm), unclonable key lengths of 225,000 bits can be extracted. Uniquely among literature, our device offers multiple methods of greatly expanding this keyspace on-the-fly, without the need to remove the chip from the setup or realign input fibres. Each of our rings has its own thermo-optic shifter, and like the ring radii, no two thermo-optic shifters are the same length. We have shown in Fig. 6 that each random voltage setting effectively produces a new virtual device, with extracted keys having weak correlations and high Hamming fractions. In the vector space of all the applied voltage configuration settings, the minimum Euclidean distance between any pair of voltage settings is approximately 0.1 V. We may conservatively use this as the minimum voltage setting increment on any single heater element needed to produce fresh keys. Over the four ring heaters, with a maximum voltage of 5 V (set in this case by hardware limitations with the multi-channel power supply), this produces a space of $6.765 \times 10^6$ distinct settings. Thus, with each individual key being 225,000 bits long, this makes our keyspace $1.522 \times 10^{12}$ bits, or approximately 1.50 Tb. The keyspace can be further expanded if we consider that the device operates in TM as well as TE, has two operational input ports, and two operational output ports. Thus the final keyspace is 12 Tb (1.5 Tb $\times$ 2 polarizations$\times$ 2 input ports$\times$ 2 output ports). We have also shown that at high powers, nonlinearities in the SiN waveguides alter the spectra. However, since these effects are more localized to high Q factor resonances, their effects on keyspace expansion are minimal, though they can still serve as device authenticators. It should be noted that given the extremely low propagation losses of SiN as a PIC material platform, it is likely that the device could be scaled up to include more rings and further enhance the spectral complexity, without a massive loss penalty.

## 8. Conclusion

In conclusion, we have demonstrated a PUF in the form of a compact silicon nitride waveguide device composed of completely standard photonic components. The conceptual simplicity of the device makes it an attractive alternative to previous integrated silicon PUFs, while still leveraging stochastic fabrication variances that give each chip its unique response. Given the low propagation losses of the silicon nitride waveguides, long-lived resonant pathways in the device exist. This in turn imparts narrow spectral features on the transmission spectrum, while the broader interferometer arrangement gives a mechanism for wider phase variations to change the spectrum. Without altering the input/output conditions, using integrated micro-heaters to reconfigure the device, the keyspace of a single PUF is in excess of 1.5 Tb. Changing the input polarization or input/output port configuration can extend this keyspace to 12 Tb. Finally, by linking two chips over fibre, a spatially distributed PUF was effectively constructed. The signals and keys sent between the two ends of the link are measurably similar, while an eavesdropper is unable to construct this key from the constituent forward and backward components.

Being fabricated in a foundry, and being composed of standard photonic toolkit components, our device is highly interesting for enhancing classical security with one-time pad encryption at scale, and in a way that is fully compatible with standard telecommunication fibre components.

**Methods**

Optical Measurement Setup

The setup shown in Fig. 2a was used to collect the transmission data from single PUFs. The fibre-coupled output of a tuneable laser source (Agilent 8164B) was first passed through an in-line polarizer, before being edge-coupled onto the chip using a polarization maintaining lensed fibre (OZ optics). A microscope objective was used to image the output waveguide facet, whose collected throughput was passed through a polarization beam splitter and onto a camera and photodetector (Thorlabs PDA20CS2) via a second beam splitter. A high-resolution oscilloscope (Agilent MSO 6104A) was used to capture the spectra of the devices as the laser was swept, and the entire experiment was controlled in LabVIEW. To keep the chip at a stable temperature, a Peltier element in reverse bias was used to heat the sample and sample stage, fixing its temperature to 33.8°C.

Key Extraction Process

Transmission spectra were first collected with a high wavelength resolution of 0.25 pm (afforded by the high-resolution oscilloscope and rapid sweep speed of the laser). The normalized logarithmic data was then rescaled between *0* and *$2^n$-1* where *n* is the number of bits used to represent each intensity level. Then, after choosing an appropriate *x* axis sampling rate *dλ*, the intensity at each sample was expressed in *n*-bit Gray code and appended to the keystream. For each number, the most significant bit (MSB) was removed, owing to its limited entropy.

**Acknowledgements**

The authors wish to acknowledge useful and insightful discussions with Xavier Porte.

**Funding**

The authors acknowledge funding from the following sources: Royal Academy of Engineering (Research Chairs and Senior Research Fellowships), Engineering and Physical Sciences Research Council (EP/V004859/1), and Innovate UK (50414).

**Supplementary Information**

1. Design and Fabrication

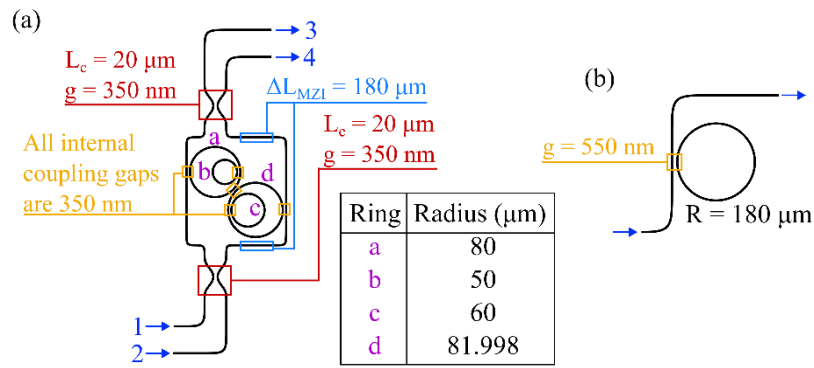

**Fig.10** (a) Schematic of the device, not showing the micro-heaters. ($L_c$ = coupling length, $g$ = coupling gap, $\Delta L_{MZI}$ = total MZI arm length difference). (b) Schematic of the comparison ring whose spectrum is shown in Fig. 2b.

A diagram of the device, with pertinent features labelled, is shown in Fig. 10a. All internal coupling gaps were set to 350 nm. All waveguides, including those of the comparison notch ring resonator fabricated on the same chip, have the same cross-sectional dimensions: 1000 nm wide by 800 nm high. Fabrication was carried out by LIGENTEC (Switzerland) as part an MPW run in their AN800 platform, which features tightly confined, single-mode operation at c-band wavelengths, and bending losses which are negligible for radii greater than or equal to 50 μm.

2. Experimental Detail – Voltage Array Settings and EDFA Measurements

Further details on the voltage setting measurements are as follows. Stability over the course of 216 such automated acquisitions was ensured by a number of methods. First, the input fibre was periodically realigned to negate the effects of any mechanical drift on the fibre-waveguide coupling. Second, a baffle was constructed around the setup to reduce unwanted airflow across the sample. Finally, the Peltier which had previously been driven in reverse bias to pin the stage temperature at a value well above room temperature was instead driven in forward bias, cooling the stage to 11.55°C, providing a stable heat sink for the global heating of the chip caused by the combined action of the ring microheaters. Finally, for the EDFA measurements, the setup was modified according to Fig. 11, but was otherwise the same as previous measurements.

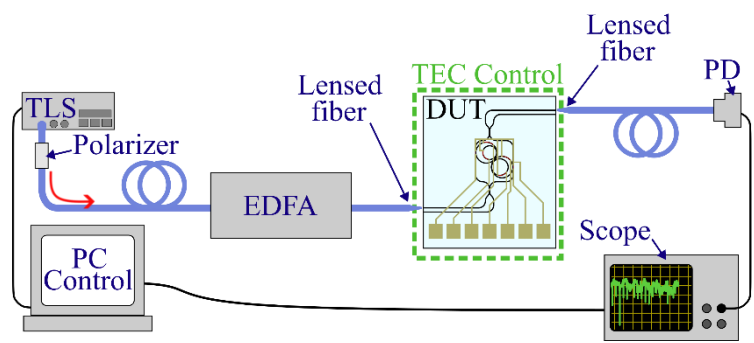

**Fig.11** (a) Setup inclusive of EDFA for measurements at high injection powers.

## 3. Polarization and I/O Port Variations

It is mentioned in the main text that both TE and TM modes are supported by the device. This can be seen in Fig. 12, which shows a representative device operating at both TE and TM. Additionally, with two input and two output ports, each device is capable of producing four possible transmission spectra. These are shown in Fig. 13. For the present work, all analysis focused on the spectra obtained from port 1 → port 3. Note in Fig. 13 that although many of the resonances occur in spectrally similar positions, their extinctions and lineshapes do vary significantly between different combinations of input/output ports.

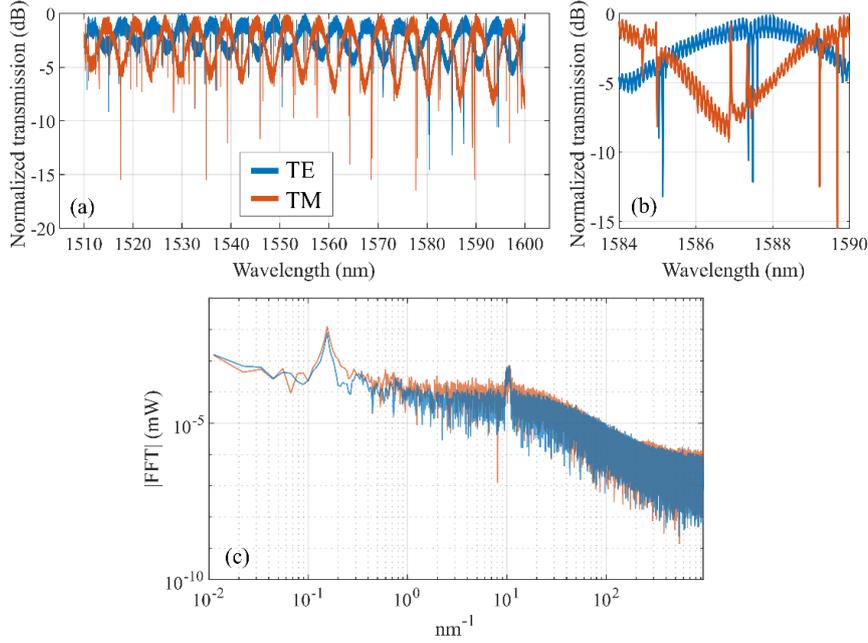

**Fig.12** (a) Comparison of the TE and TM transmission spectra taken for the same port 1 →port 2 input/output configuration. The device works for both polarizations (b) Detail of part of the same spectrum. (c) FFT of the two spectra.

## 4. Resonant Feature Characterization

Resonant features like those in Fig. 3c-e were fitted using nonlinear regression methods in MATLAB. Linewidths could then be determined from those fits. In the cases where the lineshapes deviated from a common Lorentzian, the following equation which effectively describes a Fano resonance was used [23]:

$$I(\lambda) = A + B \frac{(cot(\delta) + 2(\lambda - \lambda_R)/\gamma)}{1 + (2(\lambda - \lambda_R)/\gamma)^2}$$

The terms have the following meanings: $\lambda_R$ is the resonant wavelength, $\gamma$ is the resonant width, $A$ and $B$ are constants relating to the background signal and the height of the peak, respectively. The parameter $\delta$ relates to the Fano parameter $q$ through $q = cot(\delta)$. It describes the strength of coupling to the continuum of non-resonant modes, though in this case it effectively describes the strength of the deviation from a normal Lorentzian lineshape. It can be shown that for $q = 0$, Lorentzian lineshapes are recovered.

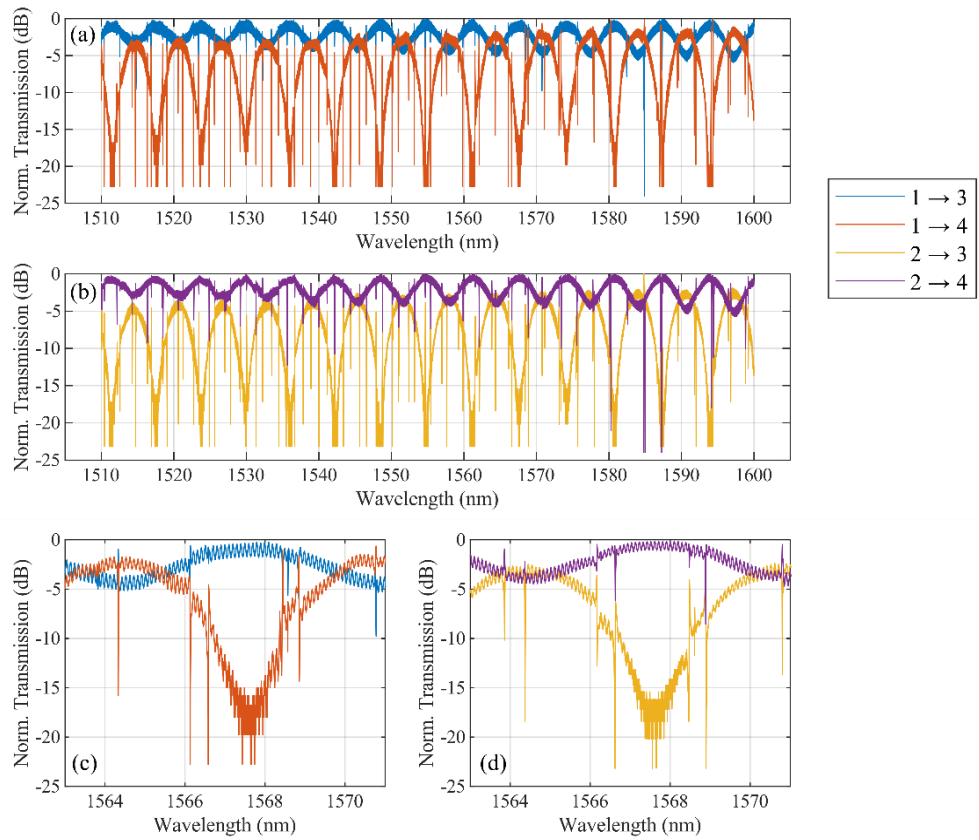

**Fig.13** (a) Comparison of the bar and cross ports for input into port 1. (b)The same for input into port 2 (c-d) Details of portions of the spectra, showing differences in both MZI background and peak extinctions.